\documentclass[amsmath]{aa}

\usepackage{amsmath}
\usepackage{txfonts}
\usepackage{natbib}
\usepackage{graphicx}

\newcommand{\uin}{u_{\rm in}}
\newcommand{\vout}{v_{\rm out}}
\newcommand{\vin}{v_{\rm in}}
\newcommand{\uout}{u_{\rm out}}
\newcommand{\ain}{a_{\rm in}}
\newcommand{\aout}{a_{\rm out}}

\newcommand{\Sigout}{\Sigma_{\rm out}}
\newcommand{\mint}{m_{\rm in}}
\newcommand{\mout}{m_{\rm out}}
\newcommand{\elik}{{\mathbf K}}
\newcommand{\md}{M_{\rm d}}
\newcommand{\psic}{\psi_{\rm c}}
\newcommand{\psit}{\psi_{\rm tot}}

\newcommand{\psir}{\tilde{\psi}_{\rm right}}
\newcommand{\psil}{\tilde{\psi}_{\rm left}}

\newcommand{\varpil}{\varpi_{\rm left}}
\newcommand{\varpir}{\varpi_{\rm right}}
\newcommand{\varpie}{\varpi_{\rm eq}}

\begin{document}

\title{A new equation for the mid-plane potential of power law disks}

\author{Jean-Marc Hur\'e\inst{1,}\inst{2} \and Franck Hersant\inst{1}}

\offprints{Jean-Marc Hur\'e}

\institute{Universit\'e Bordeaux 1
/CNRS/OASU/UMR 5804/L3AB, 2 rue de l'Observatoire, F-33270 Floirac,\\
\email{jean-marc.hure@obs.u-bordeaux1.fr \qquad franck.hersant@obs.u-bordeaux1.fr}
\and
LUTh/Observatoire de Paris-Meudon-Nancay, Place Jules Janssen, F-92195 Meudon Cedex}

\date{Received ??? / Accepted ???}

\abstract
{}
{We show that the gravitational potential $\psi$ in the plane of an axisymmetrical flat disk where the surface density varies as a power $s$ of the radius $R$ obeys an inhomogeneous first-order Ordinary Differential Equation (ODE) solvable by standard techniques.}
{The exact derivative of the midplane potentiel in its integral form is found to be algebrically linked to the potential itself.}
{The ODE reads$$\frac{d \psi}{dR} - (1+s)\frac{\psi}{R} = \Lambda(R),$$
where $\Lambda$ is fully analytical. The potential being exactly known at the origin $R=0$ for any index $s$ (and at infinity as well), the search for solutions consists of a Two-point Boundary Value Problem (TBVP) with Dirichlet conditions. The computating time is then linear with the number of grid points, instead of quadratic from direct summation methods. Complex mass distributions which can be decomposed into a mixture of power law  surface density profiles are easily accessible through the superposition principle.}
{This ODE definitively takes the place of the untractable bidimensional Poisson equation for planar calculations. It opens new horizons to investigate various aspects related to self-gravity in astrophysical disks (force calculations, stability analysis, etc.).}
{}

\keywords{Gravitation | Methods : analytical | Accretion, accretion disks}

\maketitle

\section{Introduction}

The Poisson equation plays a major role in a large variety of astrophysical problems where the dynamics of particles and fluids is influenced by gravitation. Like most Partial Differential Equations (PDEs), it is difficult to solve accurately in space and particularly inside matter due to its tridimensional nature and because it needs proper boundary conditions which are not easily accessible in the vicinity of systems. There are many situations where potential values or forces are required in a single plane where matter is gathered through a flat disk. In such cases, the Poisson equation in cylindrical coordinates $(R,\phi,z)$
$$\left(R^{-1}\partial_R R \partial_R + R^{-2}\partial^2_{\phi\phi}+ \partial^2_{zz}\right) \psi = 4 \pi G \rho(R,\phi,z),$$
where $\rho$ is the mass density and $G$ is the gravitation constant, {\it is of no use} without i) the extension of the computational grid along the $z$-direction perpendicular to the disk plane and ii) the additional knowledge of the gravity vertical gradient in the plane of the disk | the most critical point. Unfortunately, the quantity $\partial^2_{zz} \psi$ is not available in a simple manner. This is the reason why $\psi$ is always determined through direct summation techniques. 

In this letter, we report the discovery of a new local equation satisfyied by the midplane potential $\psi$ in flat axially symmetrical disks provided the surface density $\Sigma = \int_{-\infty}^\infty{\rho dz}$ varies as a power law of the cylindrical radius $R$. As shown, it is an inhomgeneous first-order Ordinary Differential Equation (ODE) subject to perfectly known Dirichlet boundary conditions. In some sense, this {new equation circumvents the $\partial^2_{zz} \psi$-problem quoted above by converting the genuine PDE into an ODE {\it for planar calculations}. Also, it is solvable by standard algorithms with very low computational cost scaling linearly with the number $N$ of grid points (instead of $N^2$ with  summation methods). As power laws are often met in astrophysical disks \citep[e.g.][]{pringle81,andwil05}, this ODE seems very well suited to investigate all these cases where disk gravity is to be acounted for. More generally, this represents a significative progress in potential theory with applications in other fields of physics.

\section{Basic theoretical considerations}
\label{sec:basicc}

The potential $\psi$ due to a flat axially symmetric disk in its plane at a cylindrical distance $R$ from the center is \citep[e.g.][]{durand64}
\begin{equation}
\psi(R) = -2G \int_{\ain}^{\aout}{\sqrt{\frac{a}{R}} \Sigma(a) m \elik(m)da},
\label{eq:psi}
\end{equation}
where $\elik$ is the complete elliptic integral of the first kind, $m$ is the modulus with
\begin{equation}
m=\frac{2\sqrt{aR}}{a+R} \quad {\rm and} \; 0 \le m \le 1,
\label{eq:mmodulus}
\end{equation}
$\ain \ge 0$ is the inner edge and $\aout > \ain$ is the outer edge. We consider surface density profiles in the disk of the form
\begin{equation}
\Sigma(a) = \Sigout \left( \frac{a}{\aout}\right)^s,
\label{eq:sigma}
\end{equation}
where $\Sigout$ is the outer edge value and $s$ is a real exponent. This choice is natural. Power laws with negative exponents are commonly met in disk theories \citep[e.g.][]{pringle81} and observations as well \citep[e.g.][]{andwil05}. These can also be combined to describe other kinds of mass distributions (see Sect. \ref{sec:complex}). The total disk mass is then
\begin{equation}
\md = 2 \pi \int_{\ain}^{\aout}{\Sigma(a) a da} = 2 \pi \Sigout \aout^2 \xi(\Delta),
\label{eq:tmass}
\end{equation}
where\footnote{\label{note:limdelta} Note that $\lim_{n \rightarrow 0} \frac{1-\Delta^n}{n} = - \ln \Delta$ for any $\Delta \ne 0$.}
\begin{equation}
\label{eq:xsi}
\xi(\Delta)=\frac{1- \Delta^{2+s}}{2+s} 
\end{equation}

A key quantity is the potential at the center $\psi(0) \equiv \psic$ which can easily be deduced from Eq.(\ref{eq:psi}) since $\elik(0) = \frac{\pi}{2}$. We have
\begin{equation}
\psic = -2 \pi G \int_{\ain}^{\aout}{\Sigma(a) da} = -2 \pi G \Sigout \aout \chi(\Delta),
\label{eq:psic}
\end{equation}
where (see note \ref{note:limdelta})
\begin{equation}
\chi(\Delta) = \frac{1 - \Delta^{1+s}}{1+s}
\label{eq:chi}
\end{equation}
and $\Delta = \ain/\aout$ is the axis ratio. Note that cases $s=\{-2,-1\}$ are not compatible with a disk filling the center (i.e.  with $\ain = 0$).

\section{Towards the ODE}

In order to obtain the ODE, we divide the $R$-axis into $3$ domains. For $R \le \ain$ (i.e. inside the central hole), we set $v=R/a \le 1$ and change the modulus of the complete elliptic integral according to the identity \citep{gradryz65}
\begin{equation}
\elik\left(\frac{2\sqrt{x}}{1+x}\right) = (1+x) \elik(x), \qquad {\rm where} \; x \le 1.
\label{eq:comod} 
\end{equation}
Since $\Sigma(a)= \Sigout \vout^s v^{-s}$, Eq.(\ref{eq:psi}) also reads
\begin{equation}
\psi(R) = 4GR \Sigout \vout^s \int_{\vin}^{\vout}{ \elik(v) \frac{dv}{v^{s+2}}},
\label{eq:psiv}
\end{equation}
where $\vin = R /\ain$ and $\vout = R/\aout$. The potential depends on five quantities: four parameters $\ain$, $\aout$, $s$ and $\Sigout$ and a single space variable $R$. The radial gradient of the potential (i.e. the opposite of the acceleration) is then given by an exact derivative 
\begin{equation}
\label{eq:drpsiv}
\frac{d \psi}{dR} = (1+s)\frac{\psi}{R} + 4GR\Sigout \vout^s \frac{d}{dR} \left\{ \int_{\vin}^{\vout}{ \elik(v) \frac{dv}{v^{s+2}}} \right\}.
\end{equation}

It is not necessary to perform the integration, as using
\begin{equation}
\frac{d}{dx} \int_{y_1(x)}^{y_2(x)}{f(y)dy} = f(y_2) \frac{d y_2}{dx} - f(y_1) \frac{d y_1}{dx},
\end{equation}
we directly get 
\begin{flalign}
\label{eq:drpsiv2}
\frac{d \psi}{dR} = & (1+s)\frac{\psi}{R} \\
&+ 4GR \Sigout \vout^s \left[\frac{\elik(\vout)}{\vout^{2+s}} \frac{d \vout}{dR} - \frac{\elik(\vin)}{\vin^{2+s}} \frac{d \vin}{dR} \right],
\nonumber
\end{flalign}
where  $dv/dR = 1/a$. Let $\tilde{\psi} = \psi/\psic$ be the normalized potential and $\varpi=R/\aout$ the normalized radial coordinate (it is equivalent to $\vout$). Using Eq.(\ref{eq:psic}) to eliminate $\Sigout$, Eq.(\ref{eq:drpsiv2}) takes the form
\begin{equation}
\frac{d \tilde{\psi}}{d \varpi}= (1+s)\frac{\tilde{\psi}}{\varpi} + S_{\rm left}(\varpi),
\label{eq:drpsivnorm}
\end{equation}
where
\begin{equation}
S_{\rm left}(\varpi) = - \frac{2}{\pi \chi(\Delta) \varpi} \left[ \elik(\varpi) - \elik\left(\frac{\varpi}{\Delta}\right)\Delta^{1+s}  \right].
\label{eq:sleft}
\end{equation}
We see that the normalized potential obeys an inhomogeneous first-order Ordinary Differential Equation (ODE), with $S_{\rm left}$ as second member and $s$ and $\Delta$ as parameters.

For $R \ge \aout$ (i.e. outside to the disk), we proceed in a similar manner with $u=a/R \le 1$. Using Eq.(\ref{eq:comod}), Eq.(\ref{eq:psi}) becomes
\begin{equation}
\psi(R) = -4GR \int_{\uin}^{\uout}{\Sigma(a) \elik(u) u du},
\label{eq:psiu}
\end{equation}
where $\uin = \ain/R$ and $\uout = \aout/R$. Taking the exact derivative of $\psi$ and using  normalized quantities, we find
\begin{equation}
\frac{d \tilde{\psi}}{d \varpi} = (1+s)\frac{\tilde{\psi}}{\varpi} + S_{\rm right}(\varpi),
\end{equation}
where
\begin{equation}
S_{\rm right}(\varpi) = - \frac{2}{\pi \chi(\Delta) \varpi^2} \left[ \elik\left(\frac{1}{\varpi}\right) - \elik\left(\frac{\Delta}{\varpi}\right)\Delta^{s+2} \right].
\end{equation}
This is the same ODE as Eq.(\ref{eq:drpsivnorm}), but with a different second member.

Finally, for $\ain \le R \le \aout$ (i.e. within the disk), we use both the variable $v$ and $u$. The potential reads
\begin{flalign}
\label{eq:psiuv}
\psi(R) =  -4GR \Sigout \vout^s & \left\{ \int_{\uin}^1{ \elik(u) u^{1+s} du} \right.\\
 & \qquad \left. -  \int_1^{\vout}{\elik(v) \frac{dv}{v^{2+s}}} \right\},
\nonumber
\end{flalign}
and so
\begin{equation}
\label{eq:drpsiuv}
\frac{d \psi}{dR} = (1+s)\frac{\psi}{R} + 4G \Sigout \left[ \frac{\elik(\vout)}{\vout}  - \elik(\uin) \uin^2 \right].
\end{equation}
Multiplying this expression by $\aout/\psic$ and using Eq.(\ref{eq:psic}) again to eliminate $\Sigout$, we obtain the expression
\begin{equation}
\frac{d \tilde{\psi}}{d \varpi} = (1+s)\frac{\tilde{\psi}}{\varpi} + S_{\rm inside}(\varpi).
\end{equation}
We recognize the same ODE as above, but with yet another second member, namely
\begin{equation}
S_{\rm inside}(\varpi)= - \frac{2}{\pi \chi(\Delta) \varpi^2} \left[ \varpi \elik(\varpi) - \elik\left(\frac{\Delta}{\varpi}\right) \Delta^{s+2}  \right].
\end{equation}

\section{Asymptotic properties}
\label{sec:asymp}

It is interesting to verify that the differential equation possesses the right properties both at small and at large distances. A second order expansion of the term $S_{\rm left}(\varpi)$ as $\varpi/\Delta \rightarrow 0$ for $\ain > 0$ shows that \citep{gradryz65}
\begin{equation}
\lim_{\varpi \rightarrow 0} \; \varpi S_{\rm left}(\varpi)  = -(1+s)-\frac{\left( 1-\Delta^{s-1} \right)}{4 \chi(\Delta)}\varpi^2 .
\end{equation}
As a consequence, we get the approximation
\begin{equation}
\varpi \frac{d (\tilde{\psi}-1)}{d \varpi}  - (1+s)(\tilde{\psi}-1) \approx - \frac{\varpi^2 \left( 1-\Delta^{s-1} \right)}{4 \chi(\Delta)},
\label{eq:odeapproxu}
\end{equation}
which admits a trivial solution, namely
\begin{equation}
\tilde{\psi}(\varpi) \approx 1 + \frac{\varpi^2}{4 \chi(\Delta)} \left(\frac{1-\Delta^{s-1}}{s-1}\right).
\label{eq:psinear0}
\end{equation}
As expected, the potential is quadratic with $\varpi$ close to the center, implying the linearity of the gravitational acceleration with $\varpi$. This approximate expression is fully compatible with Eq.(\ref{eq:psiv}) provided the complete elliptic integral is preliminarily second-order expanded over the modulus as $R \rightarrow 0$. Note that the ODE is not defined at the center. This is however not a big problem since $\tilde{\psi}=1$ and $d \tilde{\psi}/d \varpi=0$ at $\varpi=0$ by symmetry.

Conversely, for $R \gg \aout$, the potential is expected to merge with that of a central point mass with mass $\md$. We have  $m \rightarrow 0$ as $\varpi \rightarrow \infty$, hence $\chi(\Delta) \varpi^2 S_{\rm right}(\varpi) \rightarrow  \Delta^{s+2}-1$. From Eqs.(\ref{eq:tmass}) and (\ref{eq:psic}), we then find
\begin{equation}
R\frac{d \psi}{dR} \approx s \left(\psi +\frac{G \md}{R} \right) + \psi+ 2\frac{G \md}{R},
\end{equation}
which must be fulfilled for any $s$. This has a double implication: $\psi \rightarrow - \frac{G \md}{R}$ and $-d \psi/d R \rightarrow - \frac{G \md}{R^2}$ as $ R \rightarrow \infty$. This is correct.

\section{The ODE in compact form}
\label{sec:unified}

As shown, at any point in the disk plane, $\tilde{\psi}$ satisfies the ODE
\begin{equation}
\frac{d \tilde{\psi}}{d \varpi} - (1+s)\frac{\tilde{\psi}}{\varpi} = S(\varpi),
\label{eq:compactode}
\end{equation}
where the second member is a piece-wise defined function
\begin{equation}
S(\varpi) = \begin{cases}
S_{\rm left}(\varpi),  \qquad {\rm for} \quad 0 \le \varpi \le \Delta,\\
S_{\rm inside}(\varpi), \qquad {\rm for} \quad \Delta \le \varpi \le 1,\\
S_{\rm right}(\varpi), \qquad {\rm for} \quad \varpi \ge 1.
\end{cases}
\end{equation}

These three expressions are apparently different. However, they can be converted back into a single function by changing the modulus of the function $\elik$ using Eq.(\ref{eq:comod}). From Eq.(\ref{eq:mmodulus}), we have
\begin{flalign}
\label{eq:mbounds}
\mint = \frac{2 \sqrt{\Delta \varpi}}{\Delta+\varpi} , \qquad {\rm and} \qquad \mout =  \frac{2 \sqrt{\varpi}}{1+\varpi},
\end{flalign}
and so we get the expression
\begin{equation}
S(\varpi)= \frac{ \elik(\mint) \mint \Delta^{s+\frac{3}{2}} - \elik(\mout) \mout }{\pi \chi(\Delta) \sqrt{\varpi^3}}.
\label{eq:unifieds}
\end{equation}
valid for any $\varpi \in [0,\infty[$}. One thus recognizes the expression given in the abstract, with $\Lambda(\varpi) = \psic  S(\varpi) / a_2$. This ODE  can obviously be rewritten in different equivalent forms. With caution, it can even be deduced by direct calculus (see the Appendix). Note that Eq.(\ref{eq:compactode}) is compatible with the famous solution of \cite{mestel63} where the square of the disk rotational velocity $- d_{\ln \varpi} \tilde{\psi}$ is uniform for $s=-1$ provided the disk has infinite mass/extension (i.e. $\ain=0$ and $\aout=\infty$).

\section{Gravitational forces and zero-gravity point}
\label{sec:forces}

Eq.(\ref{eq:compactode}) can also be regarded as an algebraic relationship between the potential $\psi$ and the radial acceleration of gravity $g_R = -d\psi/dR$. This is particularly attractive from a dynamical point of view since the force $F_R = \mu g_R$ undergone by a particle with mass $\mu$ moving inside the disk plane is directly accessible from the potential (instead of its gradient), by the formula
\begin{equation}
F_R(R) = - \mu \frac{\psic}{\aout} \left[ \frac{(1+s)}{\varpi} \tilde{\psi}(\varpi) - S(\varpi) \right].
\end{equation}
There is always one point inside any disk where the potential goes through a maximum (and the total force vanishes). Let $\varpie \in [\Delta,1]$ be this equilibrium radius in units of $\aout$; we have
\begin{equation}
(1+s)\tilde{\psi}_{\rm eq} = - \varpie S(\varpie).
\end{equation}

\section{The ODE, in practice}
\label{sec:numode}

As it is well known, ODEs are easier to solve than PDEs. Here, $\tilde{\psi}$ is exactly known in two places in the disk plane: at the center where $\tilde{\psi}=1$ by construction, and at infinity where $\tilde{\psi}=0$ by nature. Thus, the problem is intrinsically a two-point boundary value problem (TBVP) with exactly known Dirichlet conditions. The presence of the second member $S$ presents no special difficulty since special functions can be determined at computer precision from numerical libraries. Obviously, any finite size domain $[\varpil \ge 0,\varpir < \infty]$ can also be considered, provided the normalized potential is accurately determined at the left bound $\psil = \tilde{\psi}(\varpil)$ and right bound $\psir = \tilde{\psi}(\varpir)$ of the interval. For instance, on a regular $\varpi$-grid made of $N+1$ mesh points, each with coordinate $\varpi_i = \varpil + i \ell$ where $\ell$ is the mesh size (i.e. $N\ell = \varpir-\varpil$) and $i=0,\dots,N$, then the ODE can be discretized as follows
\begin{equation}
\tilde{\psi}_{i+1} - 2(2+s) \ell \frac{\tilde{\psi}_i}{\varpi_i} - \tilde{\psi}_{i-1} = 2 \ell S_i, \qquad {\rm for} \quad i=1,N-1,
\label{eq:tbvp}
\end{equation}
at second order, with $\tilde{\psi}_0 \equiv \tilde{\psi}_{\rm left}$ and $\tilde{\psi}_N \equiv \tilde{\psi}_{\rm right}$. These $N-1$ algebraic expressions can be put in matrix form, and rapidly solved for $\tilde{\psi}_i$ by standard techniques (e.g. with the Thomas algorithm).

If a potential value $\psil$ is known at a ``starting'' radius $\varpil$ (possibly the center), solving the ODE amounts to an Initial Value Problem (IVP) since Eq.(\ref{eq:compactode}) also writes
\begin{equation}
\tilde{\psi}(\varpi) = \psil + \int_{\varpil}^{\varpi}{\left[(1+s)\frac{\tilde{\psi}(\varpi')}{\varpi'} + S (\varpi') \right] d\varpi'}.
\label{eq:ivp}
\end{equation}
A $1$st-order implicit scheme (slightly better than an explicit scheme) appropriate for $s<0$ would then give at $\varpi_{i+1}$ the value
\begin{equation}
\tilde{\psi}_{i+1} = - \frac{\varpi_{i+1} \left[ \tilde{\psi}_i + (\varpi_{i+1}-\varpi_i) S_{i+1} \right]}{s\varpi_{i+1}-(1+s)\varpi_i},
\label{eq:ivpeuler}
\end{equation}
and so the solution can be carefully propagated up to $\varpir$.

\begin{figure}
\includegraphics[width=9.0cm]{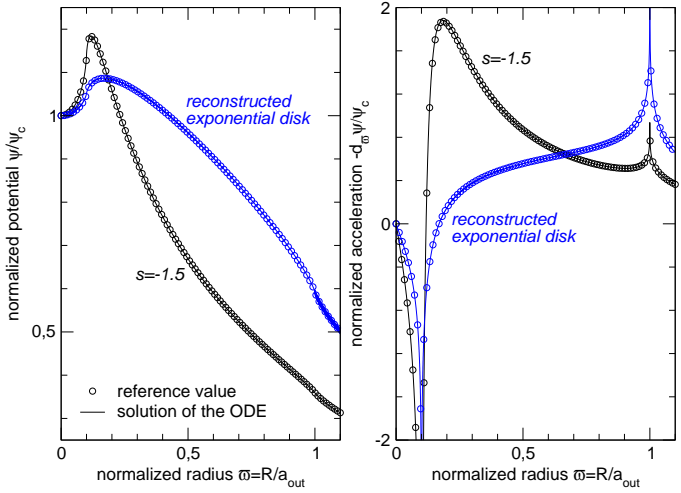}
\caption{Normalized potential $\tilde{\psi}$ obtained by solving the ODE {\it (left)}, and  acceleration of gravity deduced from Eq.(\ref{eq:compactode}) {\it (right)} (see text for the disk parameters and setup). Reference values are computed from the splitting method by \cite{hurepierens05}. The relative errors, less than $0.1 \%$, are not visible on the graphs.}
\label{fig:odepsi}
\end{figure}

We immediately see two major advantages of solving the ODE with these methods: first, the computing time is expected to be proportional to the number $N$ of grid points (instead of $N^2$ with direct summation methods) and second, the method simultaneously supplies the potential and the accceleration. Let us give a simple example. Figure \ref{fig:odepsi}a displays the potential $\tilde{\psi}_i(\varpi_i)$ when the solution of the IVP is propagated from the center $(\varpi_0,\tilde{\psi}_0) \equiv (0,1)$ through a disk using Eq.(\ref{eq:ivpeuler}) with the following parameters: $N=1000$ and $\varpi_N=1.1$ for the grid, and $\ain = 0.1$, $\aout = 1$ (axis ratio $\Delta=0.1$) and $s=-1.5$ for the disk. Figure \ref{fig:odepsi}b displays the corresponding acceleration $g_\varpi = -d\tilde{\psi}/d\varpi$ obtained from Eq.(\ref{eq:compactode}). We have compared these data with the $16$-digit reference values obtained from the splitting method by \cite{hurepierens05}. The relative errors are of the order $3 \times 10^{-4}$ on average, which is already remarkably low. Better results can be obtained with other schemes. A crude comparison of computing times shows that, at a given precision level, potential values and accelerations are determined together much more rapidly from the ODE (by a factor $\sim 40$ in the present example) than from direct summation. This is far from being negligeable, especially if the potential is to be determined many times.

\section{Complex distributions}
\label{sec:complex}

The linearity of the Poisson equation enables the description of complex distributions of matter from individual systems through the superposition principle. This is particularly attractive since power laws represent probably the most natural (if not the simplest) set of basis functions (in particular, a basis for polynomials). So, a composite system made up of $J$ co-planar and co-axial disks, each with inner edge $a_{{\rm in}, j}$, outer edge $a_{{\rm out},j}$, axis ratio $\Delta_j$, outer surface density $\Sigma_{{\rm out},j}$ and power index $s_j$, generates at distance $R$ from the center a total potential
\begin{equation}
\psit(R) = \sum_{j=1}^J{\tilde{\psi}_j(\varpi_j) {\psi_{{\rm c}_j}}},
\end{equation}
where $\varpi_j = R/a_{{\rm out},j}$ and $\psi_{{\rm c}_j} = -2 \pi G \Sigma_{{\rm out},j} a_{{\rm out},j} \chi(\Delta_j)$ according to Eq.(\ref{eq:psic}). The problem then involves $J$ independant ODEs which can be rapidly solved, possibly in parallel. This is illustrated in Fig. \ref{fig:odepsi} which shows the potential (same disk as before) for an exponential profile $\Sigma(a) \propto \exp({-a/\aout})$ reconstructed from its Taylor expansion (a series of power laws). Here, $\psit$ has been determined from the TBVP and space mapping, with $J=20$ terms (sufficient to perfectly match the exponential in the range $[0,1]$ with $16$ digits).

\section{Concluding remarks}

We have demonstrated that the classical Poisson equation can advantageously be replaced by an ODE when potential values are required in the plane of an axially symmetrical flat finite size disk where the surface density is a power law of the radius. Some properties of the ODE have been discussed and a simple example of numerical integration through one of the most basic schemes has been given.  As quoted, the combination of power law profiles make the description of a wide variety of matter distributions possible. This work can be analyzed in much more detail (like the numerical implementation of the ODE) and extended in several ways. For instance, it would be very interesting to consider other density profiles, the off-plane case as well as the full planar case by releasing the $\phi$-invariance. These questions will be tackled in forthcoming papers.

\begin{acknowledgements}
F. Hersant was supported by a CNRS fellowship which is gratefully ackonwledged. We thank J. Braine and the anonymous referee for valuable comments.
\end{acknowledgements}


\appendix

\section{Direct derivation of the ODE}

Equation (\ref{eq:compactode}) can be derived directly from Eq.(\ref{eq:psi}) by integrating the kernel over the modulus $m$. By reversing Eq.(\ref{eq:mmodulus}), we find
\begin{equation}
\frac{a}{R} = \left(\frac{1 \pm m'}{m}\right)^2 \equiv u(m),
\end{equation}
where $m' = \sqrt{1-m^2}$ is the complementary modulus. The surface density $\Sigma(a)$ is then an explicit function of $m$ and $R$ with
\begin{equation}
\Sigma(a) \equiv \Sigma(m,R) = \Sigout u^s(m) \uout^{-s},
\end{equation}
as well as the exact derivative
\begin{equation}
\left(\frac{da}{dm}\right)_R = \frac{2Ru(m)}{m}\left[\frac{1+u(m)}{1-u(m)}\right].
\end{equation}

Under these conditions, the midplane potential writes
\begin{equation}
\psi(R) = -2 G \Sigout R \uout^{-s} \int_{\mint}^{\mout}{u^{s+\frac{1}{2}}(m) \left(\frac{1}{R}\frac{da}{dm}\right) \elik(m)m dm}.
\label{eq:psigen}
\end{equation}

Note that the term $\frac{1}{R}\frac{da}{dm}$ (and subsequently the whole integrand) is a function of $m$ only. As $R\uout^{-s} \propto R^{1+s}$, we have
\begin{flalign}
\label{eq:psigen2}
\frac{d \psi}{d R} & = (1+s) \frac{\psi}{R}\\
& -2G \Sigout R \uout^{-s} \frac{d}{dR} \left\{\int_{\mint}^{\mout}{u^{s+\frac{1}{2}}(m) \left(\frac{1}{R}\frac{da}{dm}\right) \elik(m) m dm}\right\}
\nonumber
\end{flalign}

Using identity (\ref{eq:comod}), we find
\begin{flalign}
\nonumber
\frac{d \psi}{d R} &= (1+s) \frac{\psi}{R}  -2 G \Sigout \uout^{-s} \left\{ \uout^{s+\frac{1}{2}} \left(\frac{da}{dm}\right)_{\aout} \elik(\mout)\mout  \frac{d\mout}{dR}  \right. \\
& \qquad \qquad \qquad \left. - \uin^{s+\frac{1}{2}} \left(\frac{da}{dm}\right)_{\ain}  \elik(\mint) \mint  \frac{d\mint}{dR}  \right\}.
\end{flalign}

Since
\begin{equation}
\left(\frac{dm}{dR}\right)_a = \frac{m}{2R}\left[\frac{u(m)-1}{u(m)+1}\right],
\end{equation}
the above expression can be rearranged into 
\begin{flalign}
\frac{d \psi}{d R} & = (1+s) \frac{\psi}{R} \\
& + 2 G \Sigout  \uout^{\frac{3}{2}} \left[ \elik(\mout)\mout - \Delta^{s+\frac{3}{2}}  \elik(\mint) \mint\right],
\end{flalign}
which is equivalent to Eqs.(\ref{eq:compactode}) and (\ref{eq:unifieds}).

\end{document}